# Coupling hotspots: distinguishing between positive and negative land-atmosphere interaction

Jun Yin[1], Amilcare Porporato[2,3]

[1]Department of Hydrometeorology, Nanjing University of Information Science and Technology, Nanjing, China

[2]Department of Civil and Environmental Engineering, Princeton University, Princeton, USA.

[3]High Meadows Environmental Institute, Princeton University, Princeton, NJ, USA

**Abstract**

Understanding the complex interactions between land surface and atmosphere is essential to improve weather and climate predictions. Various numerical experiments have suggested that regions of strong coupling strength (hotspots) are located in the transitional climate zones. However, atmospheric processes in these hotspots are found to have different responses to the perturbation of surface properties. Here we establish analytical relationships to identify key role of soil moisture variances in controlling the coupling hotspots. Using the most recent numerical experiments, we find different signs of feedback in two such hotspots, suggesting the coupling can either reinforce or attenuate persistent extreme climates. We further uncover new coupling hotspots in regions where precipitation is highly sensitive to soil moisture perturbation. Our results highlight the importance of both signs and magnitudes of land-atmosphere interactions over extensive regions, where the ecosystems and communities are particularly vulnerable to the extreme climate events.

# 1. Introduction

By exchanging energy, water, and momentum with the atmosphere, the land surface influences weather patterns and the global water cycle, playing a crucial role in the Earth's climate system. This land-atmosphere coupling contributes to climatic persistence and enhances extreme events such as droughts, heatwaves, and floods, which in turn threaten food security, human health, socioeconomic development as well as biodiversity and ecosystem services[1–3]. Several numerical experiments and observational studies have been devoted to understanding such coupling processes, aiming to improve seasonal weather forecasts and climate projections. In particular, the pioneering Global Land-Atmosphere Coupling Experiment (GLACE) has been designed to diagnose land-atmosphere coupling strength by comparing ensembles of coupled and uncoupled simulations, in which soil moisture is either free to evolve or prescribed at each time step[4,5]. Due to its significance in understanding our climate system, GLACE has been carried out with regional climate models[6] and further implemented in the Coupled Model Intercomparison Project Phases 5 (CMIP5) models[7]. The Land Surface, Snow and Soil Moisture Model Intercomparison Project (LS3MIP), endorsed by CMIP6, provides a comprehensive assessment of land-climate feedbacks in the latest Earth System Models[8].

As a result, the transitional climate zones in the central Great Plains of North America, Sahel, and India were identified as global hotspots and then became the focal regions for in-depth analysis of the land-atmosphere coupling. However, questions have been raised on whether the corresponding feedback in these hotspots is positive or negative. For example, the transitional climate zone of the central Great Plains in North America is identified as the border between the regions of positive and negative feedback from soil moisture to precipitation[9,10], while others show that signs of the feedbacks are closely related to the background environmental conditions[11–13]. Results from another coupling hotspot of India are inconclusive, with some showing weakly dependence of precipitation on soil moisture[14] and other indicating the coexistence of both positive and negative feedback[15,16].

Numerous attempts have been made to directly quantify how certain meteorological variables respond to the perturbations in soil moisture. For example, the combination of convective triggering potential and low-level humidity index has been proposed to identify whether a wet soil condition increases the chances of deep convection[17]. The mixing diagram has been used to isolate the contribution of land surface and atmosphere to the variations of near-surface temperature and humidity at sub-daily timescale[18,19]. Statistical indicators, such as the correlation coefficients among subsurface, surface, and near-surface hydrometeorological variables, have been used to describe the chain of the coupling process[6,20,21]. Other studies have directly analyzed the rainfall rates and/or frequency conditional on the precedent soil moisture, aiming to quantify soil moisture – rainfall relationships[22–24], although the strong causality direction from atmosphere to land hydrological processes often masks such relationships[10,25].

To resolve some of the discrepancies and sharpen our understanding of land-atmosphere coupling, here we combine theoretical analysis with numerical experiments to explore the global patterns and signs of land-atmosphere interactions. We theoretically frame the problem by referring to idealized conditions and assuming an ensemble of infinite size with independent realizations. This allows us to simplify the quantification of coupling strength and work analytically to elucidate the key role of soil moisture variances in driving the previous identification coupling hotspots. We also make use of the latest LS3MIP to estimate the response of rainfall to the perturbation of prescribed soil moisture, circumventing the longstanding causality problem. As a result, we demonstrate subtle yet critical coupling behaviors in global coupling hotspots. While here we focused on soil moisture – rainfall feedback as one of the key features in land-atmosphere coupling processes, the proposed theoretical framework can be readily applied to other coupling features and could draw great interests from the broad scientific community interested in climate modeling.

**2. Global coupling hotspots explained by soil moisture variance**

Disentangling the contributions to precipitation generation due to soil moisture from the external forcing is one of the main challenges in diagnosing and interpreting the land-atmosphere coupling. In GLACE, this is done by simulating precipitation, $P_u$, during a season over a relatively large size of ensemble with prescribed soil moisture, $s_u$. The ensemble average, $\hat{P}_u$, smooths out the impacts of external forcing and becomes a function of the prescribed soil moisture. Thus, with prescribed soil moisture, the impacts of atmospheric dynamics on the land surface are removed, while the feedback from the prescribed soil moisture to the atmosphere are maintained. For this reason, it is referred to as a (one-way) uncoupled experiment in GLACE and LS3MIP (hence the 'u' subscript).

Our first goal is to revisit the definition of the coupling strength to better understand its structure and controls (see Methods for details). Ensemble simulations are usually limited to finite ensemble members with similar initial conditions and/or model configurations, often leading to underdispersed statistics[26–28]. To avoid mixing these biases with the quantification of coupling strength, we assume idealized conditions with infinite-size ensemble and independent realizations (see Methods). We also assume weak seasonal variability in the 3-month simulations. This assumption avoids the treatment of complex interaction between land-atmosphere coupling and climate seasonality[29] and generally aligns with GLACE criteria, which focus on the season of the strongest coupling in the northern hemisphere and do not provide sub-seasonal measures of coupling strength. As a result, the coupling strength is obtained as (see Methods, Eq. (7))

$$\Delta\Omega \approx \frac{\left[\dfrac{d\hat{P}_u}{ds_u}\right]^2 \mathrm{Var}(s_u)}{\mathrm{E}[\mathrm{Var}(P_u \mid s_u)] + \left[\dfrac{d\hat{P}_u}{ds_u}\right]^2 \mathrm{Var}(s_u)}, \qquad (1)$$

where E(·) and Var(·) are the mean and variance operators, and $d\hat{P}_u/ds_u$ qualifies the response of $\hat{P}_u$ to the prescribed soil moisture.

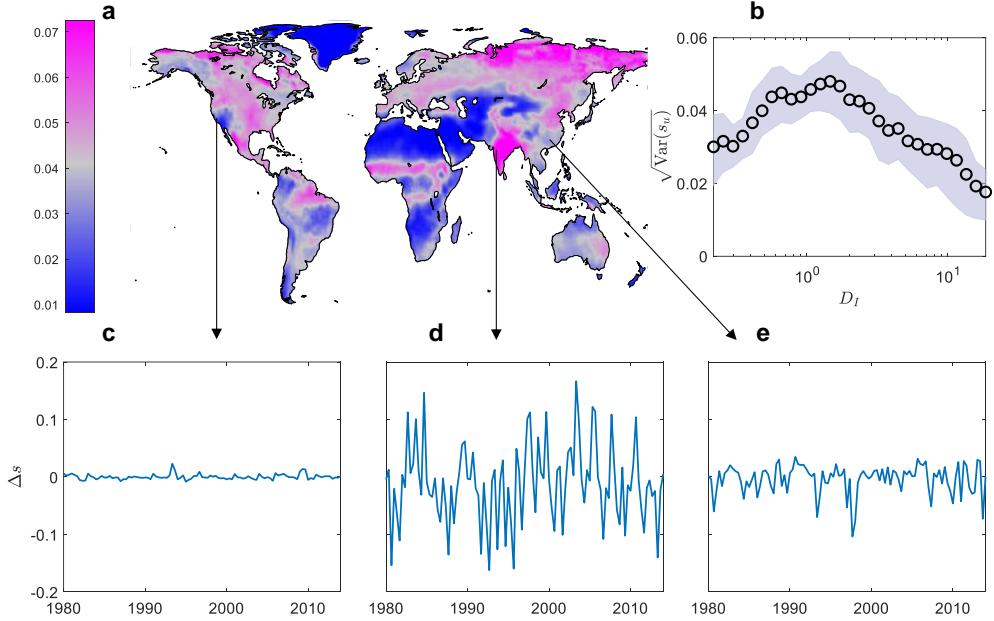

Fig. 1 Illustration of the dominant role of soil moisture variance on the global coupling strength. (a) Global distribution of ensemble average top-layer soil moisture standard deviation from multiple climate model outputs. (b) Relationships between soil moisture standard deviation and local aridity, $D_I$. The shaded area represents the first and third quartiles; the aridity was estimated as the ratio of long-term averages of potential evapotranspiration to precipitation[33]. (c-e) Time series of monthly (only in June, July, and August) top-layer soil moisture differences between historical and rmLC experiments during 1980-2014 from EC-Earth3 models in three typical climate zones of (c) California, USA, (d) Western India, and (e) Jiangxi, China. Time series from each experiment is also presented in supplementary Fig. S1. The top-layer soil moisture refers to the volumetric soil water content within the 10 cm topsoil. Climate models analyzed in this study are listed in supplementary Table S1.

Due to the square operation in Eq. (1), the signs of $d\hat{P}_u/ds_u$, indicating either positive or negative feedback, have no impact on the coupling strength. Moreover, both the square of $d\hat{P}_u/ds_u$ and the soil moisture variance, $\mathrm{Var}(s_u)$, can equally modulate the coupling strength. Thus, if the geographical variations of $d\hat{P}_u/ds_u$ are not strong, one may expect that the soil moisture variance accounts for the geographical patterns of coupling strength. To verify this point, we examine the global distributions of soil moisture variances in boreal summer from historical simulations of climate models (see Fig. 1a). As can be seen, the central Great Plains of the United States, Sahel, and India, identified as GLACE hotspots, are also high in soil moisture variance. Consistently with other studies[30–32], soil moisture variances typically increase before decreasing as regional aridity rises, leading to maximum values in climate transition zones (see Fig. 1b). In LS3MIP, the soil moisture variance is essentially the mean square deviations between the coupled historical simulations of soil moisture and the prescribed climatological values in the one-way uncoupled experiments. Three typical examples in three different climate zones are explicitly given in Fig. 1 c, d, and e. The soil moisture in the non-hotspot regions in California, USA, and Southeast China does not have strong inter-annual variability, resulting in small soil moisture differences between the historical and one-way uncoupled experiments (Fig. 1c and e). Over the Indian hotspot, the historical simulations of soil moisture have large seasonal and inter-annual variability, displaying marked contrast between the one-way uncoupled experiments and climatological soil moisture (Fig. 1d). These soil moisture differences are critical for interpreting the global land-atmosphere hotspots.

### 3. Contrasting responses of precipitation to soil moisture

Since the dominant role of the soil moisture variance in the traditional definition of coupling strength may have masked some aspects of the land-atmosphere coupling, our second goal is to refine our analysis to directly quantify the response of precipitation to the soil moisture perturbation, i.e., $d\hat{P}_u/ds_u$, including the sign of the feedback. In general, estimating $d\hat{P}_u/ds_u$ from coupled simulations or observations is challenging due to weak causal dependence of rainfall on soil moisture, which can be obfuscated by the direct influence in the opposite direction[10,25].

This causality problem can be overcome using the one-way uncoupled experiments in the latest LS3MIP, where the prescribed soil moisture is independent of atmospheric conditions. We estimated the differences in soil moisture and precipitation from two tier-1 LS3MIP experiments (pdLC and rmLC) and calculated the corresponding responses of precipitation to soil moisture as $d\hat{P}_u/ds_u \approx \Delta\hat{P}_u/\Delta s_u$ (see Methods). The results for EC-Earth3 and other models that participated in LS3MIP are reported in Fig. 2 and supplementary Figs. S2-S7. Overall, there are large inter-model differences in $d\hat{P}_u/ds_u$ with more regions showing positive signs of soil moisture–rainfall feedback (see supplementary Fig. S8). The absolute values of these precipitation responses to soil moisture are not necessarily large in the coupling hotspots identified in GLACE. For example, India and Northeast China have higher magnitudes of $d\hat{P}_u/ds_u$ in EC-Earth3, whereas parts of Europe stand out in most climate models, showing different geographical patterns than the conventional coupling hotspots.

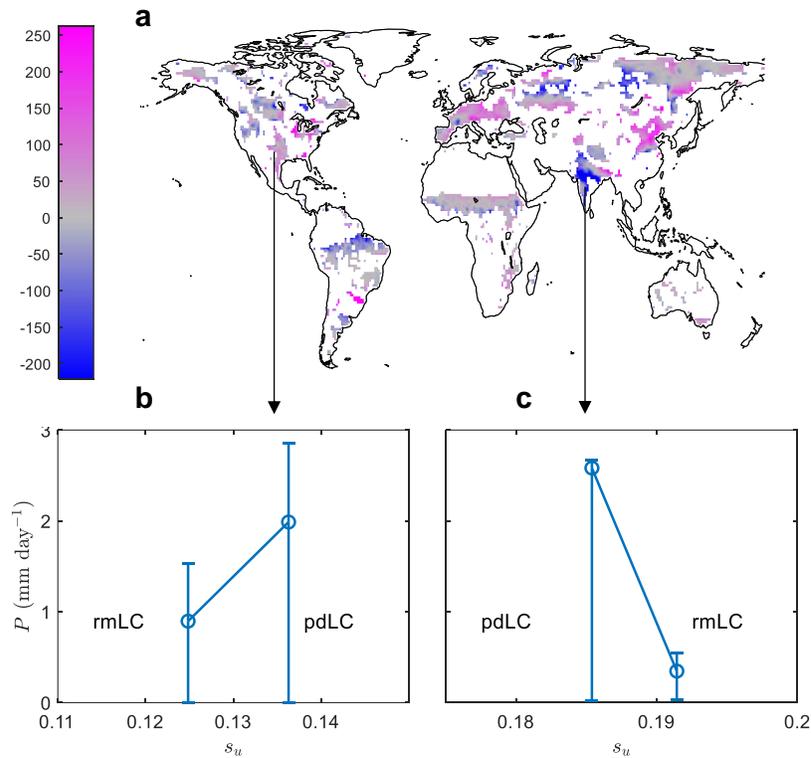

Fig. 2 Soil moisture – rainfall feedback. (a) Response of precipitation to soil moisture perturbation is estimated as the summer average precipitation differences divided by the corresponding soil moisture differences between the pdLC and rmLC experiments from EC-Earth3 model in 2014. The red and blue colors indicate positive and negative soil moisture - rainfall feedback. The white color shows regions where the soil moisture differences are too small ($|\Delta s| < 0.005$). The insets (b-c) show precipitation and soil moisture from pdLC and rmLC experiments in (b) southern Great Plain and (c) western India. The circles refer to the summer average precipitation rate; the error bars indicate the first and third quartiles of the daily precipitation rate in the summer.

Interestingly, the soil moisture – rainfall feedback tends to be negative in India and positive in the central Great Plains in North America. While both are identified as hotspots in GLACE, they show contrasting coupling behaviors with different responses of precipitation to the changes in soil moisture. For the southern Great Plains, the climatological and running mean soil moisture levels are 0.136 and 0.125 in 2014, suggesting drying trends in the last years of the historical simulations. The corresponding precipitation averaged over the boreal summer is larger under the wetter soil conditions, showing a positive feedback signal (see Fig. 2b). For India, the climatological and running mean soil moisture levels are 0.185 and 0.191 (see Fig. 2b), revealing wetting trends in recent years. The corresponding precipitation decreases from 5.3 to 0.71 mm

day$^{-1}$, respectively, corresponding to a negative feedback signal. These contrasting behaviors are significant for their impacts on understanding and modeling the regional and global water cycle.

To explain these contrasting feedbacks, we carefully examined atmospheric moisture transport and sounding profiles near these two hotspots. In the central Great Plains, the water vapor carried by the low-level jet from the Gulf of Mexico is one of the primary moisture sources for precipitation[34] (Fig. 3a). In relatively dry years, local evapotranspiration flux becomes another significant contributor to atmospheric moisture[35]. Under prescribed wetter soil conditions (pdLC experiment), more surface energy is partitioned into latent heat flux, raising low-level humidity at the cost of reducing sensible heat flux and cooling low-level air (see Fig 3 c and d). The impacts of soil moisture on the sounding profiles seem to be confined to the local scale.

In India, the southwest winds bring more abundant moisture from the Arabian Sea towards the subcontinent in the monsoon season, leading to heavy rainfall across the country (see Fig. 3b). In contrast to the central Great Plains, the local effects of surface heat flux partition in India are relatively small and may be counteracted by the remote effects on large-scale moisture advection[15]. In fact, the atmosphere is even wetter and colder under drier soil conditions (see Fig. 3 e and f). By enhancing atmospheric instability, dry soil strengthens the transport of water vapor from the ocean to the subcontinent[36–38], resulting in higher humidity around 700hPa level.

The low-level humidity could change the dynamics of atmosphere convection and may further influence regional precipitation[39–41]. Using sounding profiles averaged over the whole summer, we found that atmosphere with higher humidity in these two hotspots have lower lifting condensation levels and levels of free convection, higher levels of neutral buoyancy, larger convective available potential energy, and smaller convective inhibition (see Fig. 3 g-j), suggesting stronger buoyancy for the development of thunderstorms or severe weather. Subdaily sounding profiles, though currently not available in these experiments, may show more detailed land-atmosphere coupling processes and reveal more pronounced differences in these convective indicators.

The previous analysis clearly shows that a linkage between soil moisture, low-level humidity, moist convection, and precipitation provides an explanation of both positive and negative feedback coupling in these two major hotspots. The contrasting impacts of soil moisture on the precipitation are thus due to the cumulative outcome of both local and remote effects of land-atmosphere coupling. Wet soil tends to increase the overlying air moisture at local scale, but the reduced sensible heat fluxes from wet surface result in more stable atmosphere and possibly influence the large-scale moisture transport[42,43]. Such effects vary with location, large-scale forcing, and spatial scales of the study domain[38,44,45]. In our cases, soil moisture is prescribed at slightly different levels but approximately across the whole India and over the central Great Plains (see supplementary Fig. S9). Therefore, the soil moisture – rainfall feedback assessed here should be interpreted as subcontinental-scale coupling.

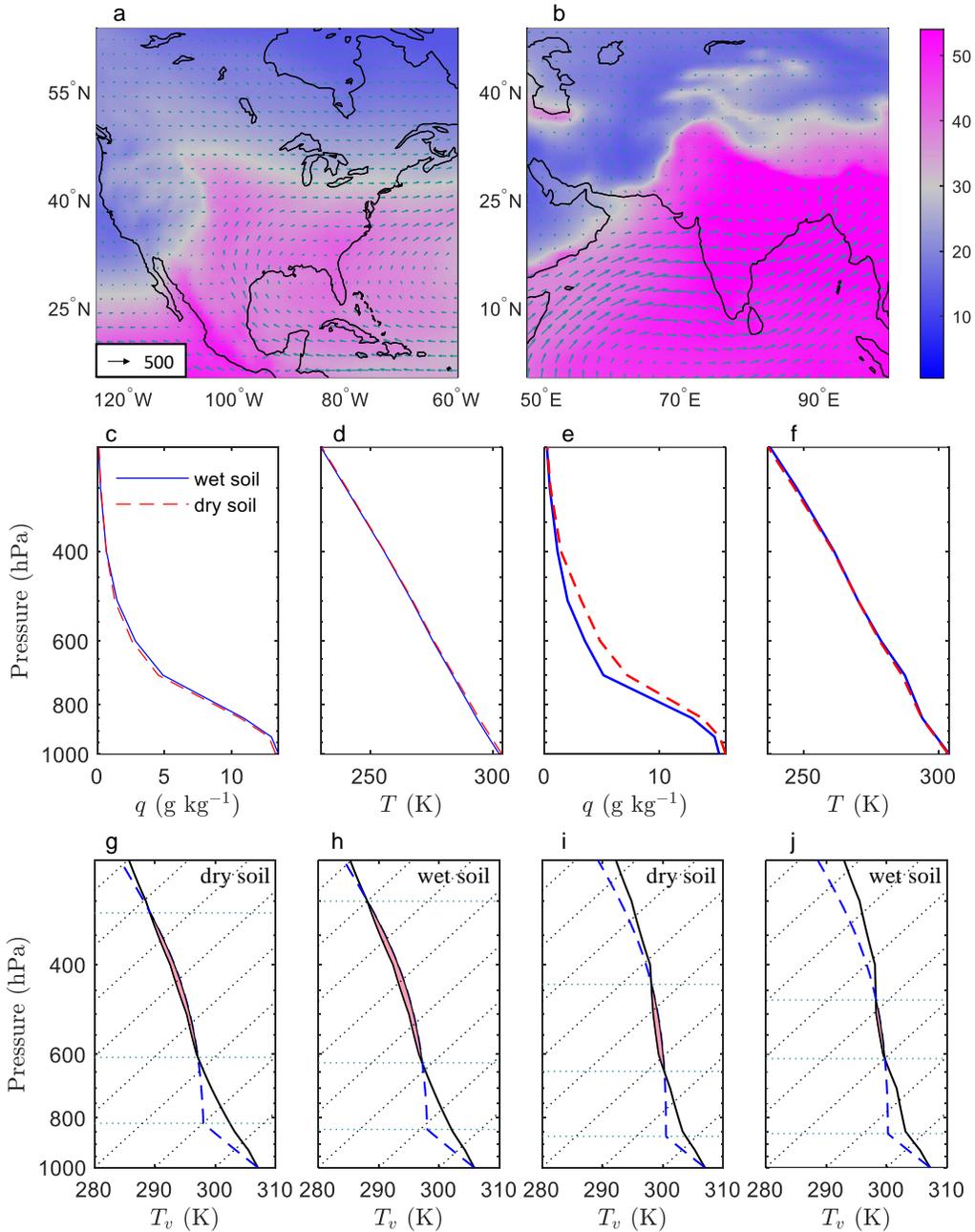

Fig. 3 Interpretation of signs of land-atmosphere coupling in central Great Plains and western India. (a, b) mean moisture transport (vectors, kg m$^{-1}$ s$^{-1}$) and vertically integrated water vapor (color maps, kg m$^{-2}$) in boreal summer in 2014 from EC-Earth3 pdLC experiment. (c-f) Sounding profiles of (c, e) specific humidity and (d, f) temperature over the wet and dry soils in (c, d) central Great Plain and (e, f) western India. (g-j) Skew-T diagrams over the (g, i) dry and (h, j) wet soils in (g, h) central Great Plain and (i, j) western India. The black lines are virtual temperature profiles, dash lines refer to moist adiabatic processes, and three horizontal dot lines indicate lifting condensation levels, level of free convection, and level of neutral buoyancy.

## 4. Conclusions

By combining theoretical considerations and careful analysis of recent climate simulations, we have demonstrated that the variations of soil moisture predominantly account for the global distributions of the coupling strength conventionally used to define land-atmosphere hotspots. Such regions in transitional climate zones are thus hydrological variability hotspots. A full analysis of the responses of precipitation to the perturbation of soil moisture shows that land-atmosphere coupling hotspots extend to some wet and dry regions.

In such dry and wet regions, positive feedback could induce either prolonged droughts or floods by creating a self-reinforcing cycle, whereas negative feedback often reduces the sensitivity of the climate system to the external forcing and stabilizes the hydrometeorological processes[38,46,47]. Therefore, the feedback signs and magnitudes are of great importance to the modeling of the persistent extreme climates, which are projected to become more frequent in some drylands and wet regions and could threaten the local ecological security and societal development[48].

From the modeling point of view, inaccurate representation of hydrological processes in such regions could result in systematical biases in modelling precipitation because of changes in soil moisture, thus triggering land-atmosphere coupling with potential for error amplifications. Consistently with the original goals of GLACE experiments, close attention to soil moisture dynamics and its representation in these new hotspots could reduce the bias propagation from soil moisture to precipitation and provide more accurate information for weather and climate projections.

Our analysis of the precipitation response to soil moisture has also important implications for managed ecosystems, where large-scale irrigation can change the soil moisture dynamics, with positive feedback enhancing precipitation and partially offsetting the need for irrigation or negative feedback suppressing precipitation and reducing the irrigation efficiency. Moreover, land use and land cover changes such as deforestation and urbanization can impact rainfall-runoff relationships and influence surface heat flux partitioning and in turn precipitation-soil moisture dynamics. Understanding the direction of the soil moisture – rainfall feedback in relation to changes in land surface properties could help diagnose how these nature-human interactions influence water resource availability.

The discriminatory power of our analysis, pertaining to sign and strength of land-atmosphere interaction, points to the need to refine the experimental designs for global assessments of the land-atmosphere coupling. Some regions could not be analyzed in this study due to the close soil moisture level in these one-way uncoupled experiments (pdLC and rmLC). New experiments with prescribed soil moisture slightly higher or lower than the climatological values would provide a more robust estimations of soil moisture – rainfall feedback at the global scale. Additional experiments with soil moisture prescribed over varying sizes of domains could be used to estimate scale-dependence of soil

moisture – rainfall feedback. These refined experiments could be immediately used to analyze other coupled variables, such as surface heat flux and near-surface air temperature, thus offering a comprehensive view of land-atmosphere coupling.

## Methods

### Soil moisture

In CMIP6, soil water is provided as the mass of water in the upper layer (100 mm) and total soil column. By dividing soil layer depth and water density, it is converted to volumetric soil water content and is referred to as soil moisture throughout the main text. In LS3MIP of CMIP6, soil moisture is prescribed in each layer of soil with climatological values; the top-layer and total-column soil moisture are spatially and temporally highly correlated (see supplementary Fig. S10-S12). For this reason, we only present the results for the top-layer soil moisture in the main text but remind the readers that similar patterns can be found for the total-column soil moisture.

### Deconstructing the coupling-strength index

In the GLACE framework, multiple sets of climate model experiments are designed to diagnose the so called coupling strength. In the coupled experiment, climate model is running with freely evolving soil moisture (subscript $c$); in one-way uncoupled experiment, soil moisture is prescribed at each time step (subscript $u$). The prescribed soil moisture in the one-way uncoupled experiment is chosen from one arbitrary realization in the coupled experiment, which is expected to reach a steady state after the initial spin-up.

Given the precipitation time series obtained from an ensemble of climate-model simulations, Koster et al. (2004) defined a coupling parameter as

$$\Omega_x = \frac{n\text{Var}(\hat{P}_x) - \text{Var}(P_x)}{(n-1)\text{Var}(P_x)}, \qquad (2)$$

where $n$ is the ensemble size, the subscript $x$ refers to either coupled experiment, $c$, or one-way uncoupled experiment, $u$, Var is variance operator, $P_x$ refers to $n$-realization of precipitation time series, and $\hat{P}_x$ is the corresponding ensemble average time series. Note that $\text{Var}(P_x)$ is precipitation variance in both the time and ensemble domains and $\text{Var}(\hat{P}_x)$ is precipitation variance only in the time domain. For $n \to \infty$, it becomes

$$\Omega_x = \frac{\text{Var}(\hat{P}_x)}{\text{Var}(P_x)}. \qquad (3)$$

The coupling strength is defined as the difference in coupling parameters between coupled and one-way uncoupled experiments

$$\Delta\Omega = \Omega_u - \Omega_c. \qquad (4)$$

With the assumption of negligible climate seasonality in the considered period (e.g., boreal summer), the stochastic characteristics of large-scale forcing and land-atmosphere coupling strength do not change. As a result, for the coupled experiments with an infinite ensemble size, the ensemble average of precipitation at any time is the climatological mean value after the initial spin-up, $\hat{P}_c(t)$=const and the variance is zero, resulting, for the coupled experiment, in $\Omega_c = 0$.

In the uncoupled experiments, the ensemble average of precipitation over an infinite ensemble size is essentially only controlled by the prescribed soil moisture, $\hat{P}_u(s_u)$. If the variance of soil moisture is known, the variance of its function, $\hat{P}_u(s_u)$, can be derived from its Taylor expansion[49,50]

$$\text{Var}\left[\hat{P}_u\right] = \left[\frac{d\hat{P}_u}{ds_u}\right]^2 \text{Var}(s_u) - \frac{1}{4}\left[\frac{d^2\hat{P}_u}{ds_u^2}\right]^2 \text{Var}(s_u)^2 + ..., \tag{5}$$

where $d\hat{P}_u / ds_u$ qualifies how $\hat{P}_u$ responds to soil moisture.

In the one-way uncoupled experiment, the distribution of precipitation rate is not only influenced by the stochastic characteristics of external forcing but also controlled by the prescribed soil moisture. The overall rainfall rate in both time and ensemble domains can be described as a compound distribution $f(P_u; s_u)$, where the prescribed soil moisture $s_u$ is independent of external forcing and can be regarded as a parameter random variable. According to the law of total variance[51,52], the variance of this compound distribution can be expressed as

$$\text{Var}(P_u) = \text{E}[\text{Var}(P_u \mid s_u)] + \text{Var}[\text{E}(P_u \mid s_u)], \tag{6}$$

where E is mean operator. This suggests that the total variance is not only the weighted average of the conditional variance (the first term on the right-hand side) but also has additional variance of the conditional means (the second term on the right-hand side). This second variance is exactly the variance of $\hat{P}_u$, i.e., $\text{Var}[\text{E}(P_u \mid s_u)] = \text{Var}(\hat{P}_u)$.

Substituting Eqs. (5) and (6) into (3) yields the coupling strength

$$\Delta\Omega = \Omega_u = \frac{\text{Var}(\hat{P}_u)}{\text{Var}(P_u)} = \frac{\text{Var}(\hat{P}_u)}{\text{E}[\text{Var}(P_u \mid s_u)] + \text{Var}(\hat{P}_u)}$$
$$\approx \frac{\left[\dfrac{d\hat{P}}{ds_u}\right]^2 \text{Var}(s_u)}{\text{E}[\text{Var}(P_u \mid s_u)] + \left[\dfrac{d\hat{P}}{ds_u}\right]^2 \text{Var}(s_u)} \tag{7}$$

where the approximate equality denotes the neglect of the higher-order terms in Eq. (5).

In GLACE, the simulations are confined to three months in boreal summer to reduce complex interactions between climate seasonality and land-atmosphere coupling. In cases where there is indeed certain climate seasonality, the variances of $\hat{P}_c(t)$ and $\hat{P}_u(t)$ in both coupled and uncoupled experiments are expected to be larger, leading to increased values in both $\Omega_u$ and $\Omega_c$. The differences between these two, defined as the coupling strength, essentially isolate the coupling strength from the climate seasonality.

**Response of precipitation to soil moisture perturbation**

To estimate the response of precipitation to soil moisture perturbation, we use the outputs from LS3MIP of CMIP6. There are two tier-1 experiments in LS3MIP, where soil moisture is prescribed at daily timescale either as climatological soil moisture (pdLC) or 30-year running means (rmLC). We compared the soil moisture and precipitation in these two experiments and estimated the response of precipitation to soil moisture perturbation as

$$\frac{d\hat{P}}{ds} \approx \frac{\hat{P}_p - \hat{P}_r}{s_p - s_r}, \tag{8}$$

where the subscripts $p$ and $r$ refer to pdLC and rmLC experiments. Since there are only one or two ensemble outputs in LS3MIP of CMIP6, we approximate ensemble averages by time averages over the 3-month boreal summer based on the ergodic hypothesis, which is usually valid in the Earth's climate system[53,54]. To have relatively large soil moisture perturbation, we focused on the last year of the historical simulation (i.e., 2014) when the soil moisture differences between these two experiments are expected to be relatively larger.

**Moisture Transport**

To quantify moisture transport, we estimate vertically integrated water vapor, $\mathcal{Q}$,

$$\mathcal{Q} = \frac{1}{g} \int_0^{p_s} q\,dp, \tag{9}$$

and vertically integrated moisture transport, $\mathcal{T}$,

$$\mathcal{T} = \frac{1}{g} \int_0^{p_s} q\mathbf{u}\,dp, \tag{10}$$

where $g$ is gravitational acceleration, $p$ is air pressure, $p_s$ is surface pressure, $q$ is specific humidity expressed as a function of pressure level, $\mathbf{u}$ is the wind vector.

The time average of moisture transport can be further decomposed as[55]

$$\overline{\mathcal{T}} = \frac{1}{g}\int_0^{p_s} \overline{q}\,\overline{\mathbf{u}}\,dp + \frac{1}{g}\int_0^{p_s} \overline{q'\mathbf{u}'}\,dp, \qquad (11)$$

where the overbars are the time averaging operators and the primes denote the low-frequency part of the variables. Since LS3MIP of CMIP6 provide variables at daily or monthly timescales, we can only estimate the mean moisture transport (i.e., the first term on the right-hand side of Eq. (11)), which are presented in Figure 3 in the main text for pdLC experiment. Additional results with the sum of both mean and low-frequency moisture transport from hourly reanalysis of ERA5 are provided in supplementary Fig. S13, which shows similar patterns as Figure 3.